\documentclass[nofootinbib,prd,aps,superscriptaddress,preprintnumbers,twocolumn,showpacs]{revtex4-1}
\usepackage[dvipdfmx]{graphicx}
\usepackage{amsmath,amssymb,amsthm,bm,bigdelim,multirow}
\usepackage{color}

\newcommand{\kslash}{k\kern-1ex /}
\newcommand{\pslash}{p\kern-1ex /}
\newcommand{\qslash}{q\kern-1ex /}
\newcommand{\lslash}{l\kern-1ex /}
\newcommand{\sslash}{s\kern-1ex /}
\newcommand{\Dslash}{D\kern-1.2ex /}

\newcommand{\beqa}{\begin{eqnarray}}
\newcommand{\eeqa}{\end{eqnarray}}
\newcommand{\be}{\[}
\newcommand{\ee}{\]}
\newcommand{\bd}{\begin{description}}
\newcommand{\ed}{\end{description}}

\newcommand{\ben}{\begin{eqnarray}}
\newcommand{\een}{\end{eqnarray}}

\def\lsim{\raise0.3ex\hbox{$<$\kern-0.75em\raise-1.1ex\hbox{$\sim$}}}
\def\gsim{\raise0.3ex\hbox{$>$\kern-0.75em\raise-1.1ex\hbox{$\sim$}}}
\def\simgt{\rlap{\lower 3.5 pt\hbox{$\mathchar \sim$}}\raise 2.0pt \hbox {$>$}}
\def\simlt{\rlap{\lower 3.5 pt\hbox{$\mathchar \sim$}}\raise 2.0pt \hbox {$<$}}

\begin{document}
\title{Utility of geometry in lattice QCD simulations}

\author{Naoya Ukita}
\affiliation{Center for Computational Sciences, University of Tsukuba, Tsukuba, Ibaraki 305-8577, Japan}

\author{Ken-Ichi Ishikawa}
\affiliation{Graduate School of Science, Hiroshima University, Higashi-Hiroshima, Hiroshima 739-8526, Japan}

\author{Yoshinobu Kuramashi}
\affiliation{Center for Computational Sciences, University of Tsukuba, Tsukuba, Ibaraki 305-8577, Japan}


\collaboration{PACS Collaboration}

\begin{abstract}
We propose a way to improve the resolution of the spatial momentum and the time interval for hadron  propagators utilizing the lattice geometry. We demonstrate the validity of the method presenting results for pseudoscalar meson energies with and without finite momenta in a large-scale quenched QCD simulation. The method should be useful especially for master-field simulations.
\end{abstract}
\date{\today}

\preprint{UTHEP-719, UTCCS-P-111, HUPD-1802}

\maketitle

\newpage
\section{Introduction}

Lattice QCD has been considered to be an ideal theoretical tool for quantitative understanding of the dynamics of the strong interaction. The main difficulties are controlling the systematic errors: quenching effects, unphysical quark masses, finite volume effects and finite lattice spacing effects. PACS-CS collaboration, which was the predecessor of PACS collaboration, got rid of the former two systematic errors by performing 2+1 flavor lattice QCD simulations with the Wilson-type quark action reducing the ud quark masses up to the physical point \cite{pacs-cs1,pacs-cs2}. After that PACS collaboration chose a strategy to make the physical volume larger at/near the physical point: We generated 2+1 flavor QCD configurations on a $(8.1\ {\rm fm})^4$ lattice at a cutoff of $a^{-1}\approx 2.3\ $GeV \cite{k-config}. Having confirmed various advantages thanks to the large volume, we have taken a further step toward even larger scale simulation \cite{pacs10}. Meanwhile, L{\"u}scher has also proposed an idea of large-scale simulation, called master-field simulation,  based on superior properties due to large volume \cite{master-field}.

In this paper, we propose a useful method, especially in the master-field (very large volume) simulations, which provides a finer resolution for the spatial momentum and the time interval in hadron propagators with the effective use of the lattice geometry. We show its validity by taking a case of the pseudoscalar meson energies with and without finite momenta on a large lattice in quenched QCD.
   
This paper is organized as follows.
In Sec.~\ref{sec:method}, we first explain how to effectively use the lattice
geometry for calculation of physical quantities in lattice QCD. 
The simulation details are given in Sec.~\ref{sec:simulation}.
In Sec.~\ref{sec:results}, we present the numerical results for the propagation of a pseudoscalar meson with and without finite momenta and their dispersion relation. Our conclusions and outlook are summarized in Sec.~\ref{sec:conclusion}.

\section{Use of lattice geometry}
\label{sec:method}

\begin{figure}
    \includegraphics[scale=0.7]{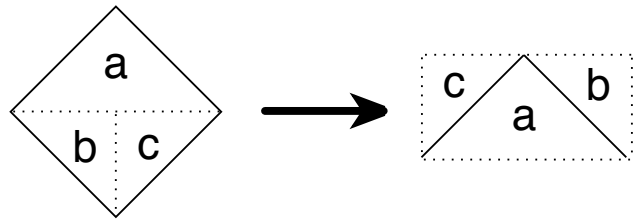}
	\caption{A reconstruction of two-dimensional periodic lattice. An $L\times L$ square lattice divided into three pieces labeled by a, b and c in the left panel can be rearranged into a rectangular lattice in the right panel with the use of periodicity. 
The long side of the rectangular lattice is periodic with the interval of $\sqrt{2}L$. The short side has the length of $L/\sqrt{2}$.    
}
    \label{fig1}
\end{figure}

We first explain a reconstruction from a given periodic square lattice to another one with different lattice geometry. Suppose we have a two-dimensional periodic square lattice with the side length of $L$, and divide it into three pieces labeled by a, b and c as illustrated in the left panel of Fig.~{\ref{fig1}}. 
By rearranging b and c pieces as in the right panel of Fig.~\ref{fig1} with the use of periodicity of the lattice, we can obtain a new rectangular lattice, whose longer side can be regarded as space with the periodic interval of $\sqrt{2}L$ 
and shorter side as time with the length of $L/\sqrt{2}$. 
 This means that the new rectangular lattice could provide the finer momentum and time resolution by $1/\sqrt{2}$ compared to the original square lattice. Note that different types of division and rearrangement for the square lattice allow different rectangular lattices. 

In order to present the rigorous procedure for the reconstruction of a four-dimensional periodic lattice, we need some mathematical preliminaries given in the following subsections. After this preparation, we will show that the reconstruction method yields lattices of different geometries with enlarged spatial volume and reduced time extent, which should improve the resolution of the spatial momentum and the discretized time interval.

\subsection{four-dimensional periodic lattice}
\label{subsec:4d}
In order to define a four-dimensional periodic lattice $\Lambda$, we start with the standard orthonormal basis  $\{{\bm e}_{\mu}\}_{\mu=1,2,3,4}$ in ${\mathbb R}^4$, 
\ben
{\bm e}_{\mu} &=& \{\left(e_\mu\right)^\alpha\equiv\delta_\mu^\alpha\}_{\alpha=1,2,3,4},\\
{\bm e}_\mu\cdot{\bm e}_\nu &\equiv& \sum_{\alpha=1}^{4} \left(e_\mu\right)^\alpha \left(e_\nu\right)^\alpha = \delta_{\mu\nu},
\een
where the first equation is the definition of the basis vectors and $\delta_\mu^\alpha, \delta_{\mu\nu}$ are the Kronecker delta and the second equation represents orthonormality of the basis. 
In this paper, we will change the orthonormal basis $\{\bm e_\mu\}$ with the Euclidian metric $\delta_{\mu\nu}$ to any basis $\{\bm E_{\mu}\}_{\mu=1,2,3,4}$ with any globally constant metric $g_{\mu\nu}$, so that a basis with lower indices and its dual basis with upper indices are different from each other in general. We distinguish upper and lower indices of vectors and tensors and use metrics to raise and lower the indices. 
Needless to say, the orthonormal basis $\{\bm e_\mu\}$ is self dual and its dual basis $\{\bm e^\mu\}$ is the same. 
Now with  the use of the orthonormal basis $\{\bm e_\mu\}$, the four-dimensional periodic lattice $\Lambda$ is defined as a set of the following coordinate vectors, 
\ben
{\bm n} &=&  n^\rho {\bm e}_\rho,\quad n^\rho\in{\mathbb Z}\label{def_e}
\een
with the Einstein summation convention.
The coefficients $\{n^\rho\}_{\rho=1,2,3,4}$ are the coordinates of ${\bm n}$ with respect to the basis $\{\bm e_\mu\}$. 
Since we assume the periodic boundary condition for all the four directions
$\{\bm e_\mu\}_{\mu=1,2,3,4}$,
we can identify $a{\bm n}$ and $a{\bm n} + aN{\bm e_\mu}$ with $a$ the lattice spacing and $N$ the lattice extent of each direction.
Note that we consider the lattice with the same lattice extent $N$ for each direction ${\bm e_\mu}$ to simplify the discussion. It is straightforward to generalize the discussion to any lattice with different lattice extents.

\subsection{Bases with reduced time and enlarged spatial volume}
Now we consider a different basis $\{\bm E_{\mu}\}_{\mu=1,2,3,4}$ 
where ${\bm E_4}$ is the time direction and $\{\bm E_i\}_{i=1,2,3}$ are the three-dimensional spatial directions perpendicular to ${\bm E_4}$. 
We first introduce the following vector ${\bm w}$ along the time direction ${\bm E_4}$, 
\ben
{\bm w} &=& w^{\rho} {\bm e_\rho},\quad w^\rho \in {\mathbb Z},\label{def_E4_tmp}
\een 
where ${\bm w}\ne{\bm e_\mu}$ and $\{w^\mu\}_{\mu=1,2,3,4}$ has no common integral divisor larger than one. 
It is worth emphasizing that the number of independent vectors  is of order of $N^3$ by a naive counting. 
For each vector ${\bm n}\in\Lambda$, we assign an integer $n_t$ in unit of the lattice spacing $a$ to express the time coordinate: 
\ben
n_t = {\rm mod}({\bm n}\cdot{\bm w},N)  =  {\rm mod}(n^\rho w^{\rho},N).
\label{def_nt}
\een
In case of ${\bm w}={\bm e_4}$, we have $n_t=n^4$ as usual. 
It allows us to divide the lattice $\Lambda$ into $N$ time slices ${B_{n_t}}$,
\ben
\Lambda &=& \bigcup_{n_t=0,1,\cdots,N-1} B_{n_t},\\
B_{n_t} &=& \{{\bm n} \in \Lambda \,|\, n_t={\rm mod}({\bm n}\cdot{\bm w},N)\}.\label{Bt}
\een
Each time slice $B_{n_t}$ contains $N^3$ lattice sites, which is explained from the periodicity on the time slices in Sec. \ref{subsec:period}. 
The distance between the adjacent time slices becomes $a/\sqrt{\bm w\cdot\bm w}< a$ with the lattice spacing $a$ divided by the norm of ${\bm w}$. So it is natural to adopt the rescaled vector ${\bm w}/({\bm w\cdot\bm w})$ as the time direction basis ${\bm E_4}$,  
\ben
{\bm E_{4}} &=& \frac{\bm w}{\bm w\cdot\bm w}\label{def_E4}\\
            &=& \left(E_{4}\right)^{\rho}{\bm e_\rho},\quad
     \left(E_{4}\right)^{\rho} =\frac{w^{\rho}}{\bm w\cdot\bm w}\in{\mathbb Q}.
\een
In fact, ${\bm E_4}$ is geometrically equivalent to the dual basis vector of ${\bm w}$ such that ${\bm E_4}\cdot{\bm w}=1$. 
Notice that the lattice spacing of the time direction, denoted by $a_4$, is reduced by the factor $\sqrt{\bm E_4\cdot\bm E_4}\ (<1)$ compared with that of the orthogonal basis,
\ben
a_4 &=& \sqrt{\bm E_4\cdot\bm E_4}\ a.
\een
So the lattice extent of the time direction is reduced to $aN\sqrt{\bm E_4\cdot\bm E_4}$ in the basis $\{\bm E_{\mu}\}$. 
The lattice volume $(aN)^4$ does not depend on the choice of basis vectors, so that 
the spatial volume should increase to $(aN)^3/\sqrt{\bm E_4\cdot\bm E_4}$ to compensate for the reduction of the time extent. In order to do so, we can take the spatial basis vectors $\{\bm E_{i}\}_{i=1,2,3}$ orthogonal to ${\bm E_4}$ such that
\ben
&&{\bm E_{i}} = \left(E_{i}\right)^{\rho} {\bm e_\rho},\quad \left(E_{i}\right)^{\rho}\in{\mathbb Z},\quad i=1,2,3,\label{def_E123}\\
&&{\bm E_{4}}\cdot{\bm E_{i}} = \left(E_{4}\right)^\alpha \left(E_{i}\right)^\alpha = 0, \quad i=1,2,3,\label{orth_E}\\
&&\det{\bm E} =\pm1,\label{detE}
\een 
where ${\bm E}$ is the $4\times 4$ matrix whose $\mu$-$\alpha$ element is $\left(E_{\mu}\right)^\alpha\ (\mu, \alpha =1,2,3,4)$. 
From purely algebraic considerations, it is found that the spatial basis vectors are uniquely fixed up to $SL(3, {\mathbb Z})$ transformations on those vectors and each of $\{\left(E_{i}\right)^\alpha\}$ for $i=1,2,3$ has no common integral divisor larger than one. 
The third equation (\ref{detE}) restricts the transformation from $\{\bm e_\mu\}$ to $\{\bm E_{\mu}\}$ to the group of linear transformations preserving the volume so that the lattice volume is $(aN)^4$ in both bases. 

\subsection{Dual basis and metric}
For later convenience we introduce the dual basis $\{\bm E^{\mu}\}$ with upper indices as the basis of momentum space, the metric $g_{\mu\nu}$ and its inverse $g^{\mu\nu}$.  
Let $\{\bm e^\mu\}$ be the dual basis of the orthonormal basis $\{\bm e_\mu\}$ such that ${\bm e^\mu}\cdot{\bm e_\nu}=\delta^{\mu}_{\nu}$ and ${\bm e^\mu}\cdot{\bm e^\nu}=\delta^{\mu\nu}$.
The dual basis $\{\bm E^{\mu}\}$ is defined as 
\ben
&&{\bm E^{\mu}}=\left(E^{\mu}\right)_{\rho}{\bm e^\rho},\label{def_dualE}\\
&&{\bm E^{\mu}}\cdot{\bm E_{\nu}}= \left(E^{\mu}\right)_{\alpha}\left(E_{\nu}\right)^{\alpha}=\delta^{\mu}_{\nu},\\
&&\left(E^{\mu}\right)_{\alpha}\left(E_{\mu}\right)^{\beta}=\delta^{\beta}_{\alpha},
\een
where $\left(E^{4}\right)_{\alpha}\in {\mathbb Z}$ and $(E^{i})_{\alpha}\in {\mathbb Q}$ for $i=1,2,3$.  
The metric and its inverse are defined as
\ben
g_{\mu\nu}&=&{\bm E_{\mu}}\cdot{\bm E_{\nu}} = \left(E_{\mu}\right)^{\alpha}\left(E_{\nu}\right)^{\alpha},\label{g}\\
g^{\mu\nu}&=&{\bm E^{\mu}}\cdot{\bm E^{\nu}} = \left(E^{\mu}\right)_{\alpha}\left(E^{\nu}\right)_{\alpha}.\label{ginv}
\een
By definition, $g_{i4}=0$ and $g^{i4}=0$ for $i=1,2,3$.  
Raising and lowering the indices are performed by contracting with the metrics as
\ben
 {\bm E^{\mu}}&=& g^{\mu\nu}{\bm E_{\nu}},\\ 
 {\bm E_{\mu}}&=& g_{\mu\nu}{\bm E^{\nu}}
\een
and equivalently
\ben
 \left(E^{\mu}\right)_{\alpha}&=& g^{\mu\nu}\left(E_{\nu}\right)^{\beta}\delta_{\beta\alpha},\\
 \left(E_{\mu}\right)^{\alpha}&=& g_{\mu\nu}\left(E^{\nu}\right)_{\beta}\delta^{\beta\alpha}.
\een

\subsection{Coordinate vector and its periodicity}
\label{subsec:period}
Having defined the geometry of $\Lambda$ in terms of the bases, any vector ${\bm n}\in \Lambda$ is expanded in both of the bases $\{\bm e_\mu\}$ and $\{\bm E_{\mu}\}$ as 
\ben
{\bm n} =  n^{\rho}{\bm e_\rho} =  r^{\rho}{\bm E_{\rho}},
\een 
where $n^\rho\in{\mathbb Z}$. 
The relation between $\{n^\mu\}$ and $\{r^{\mu}\}$ coordinates is easily derived as 
\ben
r^{\mu}=\left(E^\mu\right)_{\rho}n^{\rho},\quad   n^{\mu}=\left(E_{\rho}\right)^{\mu}r^{\rho},
\een 
and $r^4\in{\mathbb Z}$ and $r^{1,2,3}\in{\mathbb Q}$ because $\left(E^{4}\right)_{\alpha}\in {\mathbb Z}$ and $\left(E^{1,2,3}\right)_{\alpha}\in {\mathbb Q}$.  
In general, the spatial coordinates $r^{1,2,3}$ are rational numbers. 
It follows from the definition of $n_t$ in Eq.~(\ref{def_nt}) that $n_t=n^4$ in the basis $\{\bm e_\mu\}$ and $n_t=r^4$ in the basis $\{\bm E_{\mu}\}$. 

There is a remark about the periodicity of the vector $a{\bm n}$. 
In the basis $\{\bm e_\mu\}$, the vector $a{\bm n}$ is periodic with the interval of $aN{\bm e_\mu}$ for $\mu=1,2,3,4$ discussed in Sec.~\ref{subsec:4d}.  
In the basis $\{\bm E_{\mu}\}$, the spatial basis vectors in Eq.~(\ref{def_E123}) are defined by the liner combinations with the integer coefficients $\{\left(E_{i}\right)^{\alpha}\}$ for $i=1,2,3$, each of which has no common integral divisor larger than one, so that 
 $a{\bm n}$ is also periodic with the interval of $aN{\bm E_{i}}$ for $i=1,2,3$.
The spatial periodicity shows that $N$ lattice sites are aligned along each spatial direction and $N^3$ lattice sites reside in each time slice.   
On the other hand, the time direction ${\bm E_{4}}$ of Eq.~(\ref{def_E4}) is defined by 
the liner combination with the rational coefficients $\left(E_{4}\right)^{\alpha}\in{\mathbb Q}$ so that 
 $a{\bm n}$ is not periodic with the interval of $aN{\bm E_{4}}$ but periodic with  $aN{\bm E_{4}}/({\bm E_4}\cdot{\bm E_4})$.
Therefore, the time slices at $n_t(=r^{4})=0$ and $N$, given by $B_{n_t=0}$ and $B_{n_t=N}$ of Eq.~(\ref{Bt}), are not simply related by the translation along ${\bm E_{4}}$. 
We need a spatial shift vector $aN\sum_{i=1}^{3}s^{i}{\bm E_{i}}$ on $B_{n_t=N}$ to satisfy the periodicity. 
In order to determine the coefficients $\{s^i\}$, it may be convenient to introduce a vector
${\bm l}=Nl^{\rho}{\bm e_{\rho}},\ l^{\rho}\in{\mathbb Z}$ to satisfy
\ben
\frac{aN}{{\bm E_4}\cdot{\bm E_4}}{\bm E_{4}}+a{\bm l}=aN{\bm E_{4}}+aN\sum_{i=1}^{3}s^{i}{\bm E_{i}},
\een
which leads to the relation:
\ben
s^{i} = \left(E^{i}\right)_{\rho}l^\rho\in{\mathbb Q}\quad i=1,2,3.\label{smu}
\een 
For any ${\bm n}\in B_{n_t=0}$, the corresponding vector of $B_{n_t=N}$ is identified by the shift of $aN{\bm E_4}$ followed by the spatial correction of $aN\sum_{i=1}^{3}s^{i}{\bm E_{i}}$
on $B_{n_t=N}$.\footnote{This identification in ${\bm E_{4}}$ direction  corresponds to shifted boundary conditions in time direction in thermal field theories \cite{SBC}.}  The additional shift on $B_{n_t=N}$ plays an important role in time correlators with finite momenta in Sec.~\ref{t-corr}. 
  
\subsection{Improvement of resolution for time interval and spatial momentum}
Let us discuss the lattice spacing with the general basis $\{\bm E_{\mu}\}$. The lattice spacing $a_{\mu}$ for each ${\bm E_{\mu}}$ direction is defined by
\ben
a_{\mu} &=& \sqrt{g_{\mu\mu}}\ a,\quad \mu=1,2,3,4.\label{a_mu}\label{def_a_mu}
\een 
Since $0<g_{44}<1$ and $g_{11},g_{22},g_{33}\ge1$ by the definition of  Eq.~(\ref{g}), we obtain (i) $a_4< a$ and (ii) $a_{1,2,3}\ge a$. The result (i) means the improvement of the resolution for the time interval along ${\bm E_4}$ direction. The result (ii) combined with the fact that $\det(g_{\mu\nu})=1$ and $g_{i4}=0$ for $i=1,2,3$ tells us that   
 the spatial volume increases by the factor of $1/\sqrt{g_{44}}$ such as $(aN)^3/\sqrt{g_{44}}$. According to this the smallest spatial momentum squared is allowed to decrease to a value less than $(2\pi/aN)^2$. 

We now consider the Fourier series expansion to look into more details of the spatial momentum. Let $\vec{n}$ be any spatial vector given by
\ben
 \vec{n} = \sum_{i=1}^{3} r^{i}{\bm E_{i}},\quad r^{i}\in{\mathbb Q}
\een
 on the time slice $B_{n_t}$, and let $\vec{p}$ be any spatial momentum given by 
\ben
\vec{p}=\sum_{i=1}^{3}p_{i}{\bm E^{i}}
\een
 in the dual spatial basis $\{\bm E^{i}\}_{i=1,2,3}$. 
 In order to determine the allowed range of the momentum, we consider a periodic function $f(a\vec{n})=f(a\vec{n}+aN{\bm E_{i}})$ $(i=1,2,3)$ so that the Fourier series expansion is defined by 
\ben
f(a\vec{n}) = \sum_{\vec{p}} \tilde{f}(\vec{p})\exp(ia\vec{n}\cdot\vec{p}),
\een
where $\tilde{f}(\vec{p})\in {\mathbb C}$ and $\vec{n}\cdot\vec{p}=\sum_{i=1}^{3}r^{i}p_{i}$. The periodicity condition on $f(a\vec{n})$ yields the condition $\exp(iaNp_i)=1$ so that the allowed range of the momentum is 
\ben
p_i=\frac{2\pi m_i}{aN},\quad m_i=0,1,\cdots, N-1,
\een
and the spatial momentum squared is 
\ben
p^2=\vec{p}\cdot{\vec{p}}=\sum_{i=1}^{3}p_i p_j g^{ij}.\label{p}
\een
Recall that the spatial volume in the basis $\{\bm E_{\mu}\}$ increases by $1/\sqrt{g_{44}}$, and conversely its dual spatial volume in the basis $\{\bm E^{\mu}\}$ decreases by $\sqrt{g_{44}}$. 
This means that there exists  at least one diagonal spatial element of the inverse metric to satisfy  $g^{ii}<1$, where the index $i$ is not summed, in the spatial basis vectors $\{\bm E_{i}\}_{i=1,2,3}$ and their dual ones $\{\bm E^{i}\}_{i=1,2,3}$, which are, if necessary, transformed by an element of $SL(3, {\mathbb Z})$ and  its inverse respectively. Therefore, the smallest nonzero momentum squared is given by $(2\pi/aN)^2 g^{ii}$, which is smaller than $(2\pi/aN)^2$.  

\subsection{Examples}
It may be instructive to consider some representative cases for the time direction vector ${\bm w}$ of Eq.~(\ref{def_E4_tmp}).
In the following we list the relevant quantities associated with each ${\bm w}$: Basis $\{\bm E_{\mu}\}$ of Eqs.~(\ref{def_E4}), (\ref{def_E123}), its dual basis $\{\bm E^{\mu}\}$ of Eq.~(\ref{def_dualE}), the metric $g_{\mu\nu}$, $g^{\mu\nu}$ of Eqs.~(\ref{g}), (\ref{ginv}), the distance between the adjacent time slices $a_4$ of Eq.~(\ref{def_a_mu}),  the smallest nonzero momentum squared $p^2_{\rm min}$ of Eq.~(\ref{p})  and the shift coefficients $s^{i}$ in Eq.~(\ref{smu}).

\vspace{5mm}
\hspace{12mm}{(i) $\{w^1,w^2,w^3,w^4\}=\{0,0,0,1\}$}
 
The corresponding bases are the orthonormal ones: 
\ben
\{\bm E_{\mu}\}=\{\bm e_\mu\},\\ 
\{\bm E^{\mu}\}=\{\bm e^\mu\}.
\een
The metric and its inverse are the Kronecker delta $g_{\mu\nu}=\delta_{\mu\nu},\, g^{\mu\nu}=\delta^{\mu\nu}$. 
The distance between the adjacent time slices is $a_4=a$. 
The smallest nonzero spatial momentum squared is $p^2_{\rm min}=(2\pi/aN)^2$. 
The shift coefficients are $\{s^1,s^2,s^3\}=\{0,0,0\}$.

\vspace{5mm}
\hspace{12mm}{(ii) $\{w^1,w^2,w^3,w^4\}=\{0,0,1,1\}$}

The corresponding bases are
\ben
  \{\left( E_{1}\right)^1,\left( E_{1}\right)^2,\left( E_{1}\right)^3,\left( E_{1}\right)^4\}&=&\{1,0,0,0\}, \\
  \{\left( E_{2}\right)^1,\left( E_{2}\right)^2,\left( E_{2}\right)^3,\left( E_{2}\right)^4\}&=&\{0,1,0,0\}, \\
  \{\left( E_{3}\right)^1,\left( E_{3}\right)^2,\left( E_{3}\right)^3,\left( E_{3}\right)^4\}&=&\{0,0,1,-1\}, \\
  \{\left( E_{4}\right)^1,\left( E_{4}\right)^2,\left( E_{4}\right)^3,\left( E_{4}\right)^4\}&=&\left\{0,0,\frac{1}{2},\frac{1}{2}\right\},   
\een
and 
\ben
  \{\left( E^{1}\right)_1,\left( E^{1}\right)_2,\left( E^{1}\right)_3,\left( E^{1}\right)_4\}&=&\{1,0,0,0\}, \\
  \{\left( E^{2}\right)_1,\left( E^{2}\right)_2,\left( E^{2}\right)_3,\left( E^{2}\right)_4\}&=&\{0,1,0,0\}, \\
  \{\left( E^{3}\right)_1,\left( E^{3}\right)_2,\left( E^{3}\right)_3,\left( E^{3}\right)_4\}&=&\left\{0,0,\frac{1}{2},-\frac{1}{2}\right\}, \\
  \{\left( E^{4}\right)_1,\left( E^{4}\right)_2,\left( E^{4}\right)_3,\left( E^{4}\right)_4\}&=&\{0,0,1,1\}.  
\een 
The metric and its inverse are
\ben
  g_{\mu\nu}&=&\left(\begin{array}{cccc}1 & 0 & 0 & 0 \\0 & 1 & 0 & 0 \\0 & 0 & 2 & 0 \\0 & 0 & 0 & \frac{1}{2}\end{array}\right),\\  
  g^{\mu\nu}&=&\left(\begin{array}{cccc}1 & 0 & 0 & 0 \\0 & 1 & 0 & 0 \\0 & 0 & \frac{1}{2} & 0 \\0 & 0 & 0 & 2\end{array}\right).  
\een
Other relevant quantities are
\ben
 a_4&=&\frac{1}{\sqrt{2}}a,\\  
 p^2_{\rm min}&=&\frac{1}{2}\left(\frac{2\pi}{aN}\right)^2,\\
 \{s^1,s^2,s^3\}&=&\left\{0,0,\frac{1}{2}\right\}.
\een

\vspace{5mm}
\hspace{12mm}{(iii) $\{w^1,w^2,w^3,w^4\}=\{0,1,1,1\}$}

The corresponding bases are
\ben
  \{\left( E_{1}\right)^1,\left( E_{1}\right)^2,\left( E_{1}\right)^3,\left( E_{1}\right)^4\}&=&\{1,0,0,0\}, \\
  \{\left( E_{2}\right)^1,\left( E_{2}\right)^2,\left( E_{2}\right)^3,\left( E_{2}\right)^4\}&=&\{0,1,-1,0\}, \\
  \{\left( E_{3}\right)^1,\left( E_{3}\right)^2,\left( E_{3}\right)^3,\left( E_{3}\right)^4\}&=&\{0,0,1,-1\}, \\
  \{\left( E_{4}\right)^1,\left( E_{4}\right)^2,\left( E_{4}\right)^3,\left( E_{4}\right)^4\}&=&\left\{0,\frac{1}{3},\frac{1}{3},\frac{1}{3}\right\}, 
\een
and 
\ben
  \{\left( E^{1}\right)_1,\left( E^{1}\right)_2,\left( E^{1}\right)_3,\left( E^{1}\right)_4\}&=&\{1,0,0,0\}, \\
  \{\left( E^{2}\right)_1,\left( E^{2}\right)_2,\left( E^{2}\right)_3,\left( E^{2}\right)_4\}&=&\left\{0,\frac{2}{3},-\frac{1}{3},-\frac{1}{3}\right\}, \\
  \{\left( E^{3}\right)_1,\left( E^{3}\right)_2,\left( E^{3}\right)_3,\left( E^{3}\right)_4\}&=&\left\{0,\frac{1}{3},\frac{1}{3},-\frac{2}{3}\right\}, \\
  \{\left( E^{4}\right)_1,\left( E^{4}\right)_2,\left( E^{4}\right)_3,\left( E^{4}\right)_4\}&=&\{0,1,1,1\}.  
\een 
The metric and its inverse are
\ben
  g_{\mu\nu}&=&\left(\begin{array}{cccc}1 & 0 & 0 & 0 \\0 & 2 & -1 & 0 \\0 & -1 & 2 & 0 \\0 & 0 & 0 & \frac{1}{3}\end{array}\right),\\
  g^{\mu\nu}&=&\left(\begin{array}{cccc}1 & 0 & 0 & 0 \\0 & \frac{2}{3} & \frac{1}{3} & 0 \\0 & \frac{1}{3} & \frac{2}{3} & 0 \\0 & 0 & 0 & 3\end{array}\right).  
\een
Other relevant quantities are
\ben
a_4&=&\frac{1}{\sqrt{3}}a,\\
p^2_{\rm min}&=&\frac{2}{3}\left(\frac{2\pi}{aN}\right)^2,\\
\{s^1,s^2,s^3\}&=&\left\{0,\frac{2}{3},\frac{1}{3}\right\}.
\een

\vspace{5mm}
\hspace{12mm}{(iv) $\{w^1,w^2,w^3,w^4\}=\{1,1,1,1\}$}

The corresponding bases are
\ben
  \{\left( E_{1}\right)^1,\left( E_{1}\right)^2,\left( E_{1}\right)^3,\left( E_{1}\right)^4\}&=&\{1,-1,0,0\}, \\
  \{\left( E_{2}\right)^1,\left( E_{2}\right)^2,\left( E_{2}\right)^3,\left( E_{2}\right)^4\}&=&\{0,1,-1,0\}, \\
  \{\left( E_{3}\right)^1,\left( E_{3}\right)^2,\left( E_{3}\right)^3,\left( E_{3}\right)^4\}&=&\{0,0,1,-1\}, \\
  \{\left( E_{4}\right)^1,\left( E_{4}\right)^2,\left( E_{4}\right)^3,\left( E_{4}\right)^4\}&=&\left\{\frac{1}{4},\frac{1}{4},\frac{1}{4},\frac{1}{4}\right\}, 
\een
and 
\ben
  \{\left( E^{1}\right)_1,\left( E^{1}\right)_2,\left( E^{1}\right)_3,\left( E^{1}\right)_4\}&=&\left\{\frac{3}{4},-\frac{1}{4},-\frac{1}{4},-\frac{1}{4}\right\}, \\
  \{\left( E^{2}\right)_1,\left( E^{2}\right)_2,\left( E^{2}\right)_3,\left( E^{2}\right)_4\}&=&\left\{\frac{1}{2},\frac{1}{2},-\frac{1}{2},-\frac{1}{2}\right\}, \\
  \{\left( E^{3}\right)_1,\left( E^{3}\right)_2,\left( E^{3}\right)_3,\left( E^{3}\right)_4\}&=&\left\{\frac{1}{4},\frac{1}{4},\frac{1}{4},-\frac{3}{4}\right\}, \\
  \{\left( E^{4}\right)_1,\left( E^{4}\right)_2,\left( E^{4}\right)_3,\left( E^{4}\right)_4\}&=&\{1,1,1,1\}.
\een 
The metric and its inverse are
\ben
  g_{\mu\nu}&=&\left(\begin{array}{cccc}2 & -1 & 0 & 0 \\-1 & 2 & -1 & 0 \\0 & -1 & 2 & 0 \\0 & 0 & 0 & \frac{1}{4}\end{array}\right),\\
  g^{\mu\nu}&=&\left(\begin{array}{cccc}\frac{3}{4} & \frac{1}{2} & \frac{1}{4} & 0 \\ \frac{1}{2} & 1 & \frac{1}{2} & 0 \\ \frac{1}{4} & \frac{1}{2} & \frac{3}{4} & 0 \\0 & 0 & 0 & 4\end{array}\right).  
\een
Other relevant quantities are
\ben
a_4&=&\frac{1}{2}a,\\
p^2_{\rm min}&=&\frac{3}{4}\left(\frac{2\pi}{aN}\right)^2,\\
\{s^1,s^2,s^3\}&=&\left\{\frac{3}{4},\frac{1}{2},\frac{1}{4}\right\}.
\een

There is a comment on reflection positivity in terms of the  lattice site arrangement on the time slices. For each time direction ${\bm w}$ in the above examples (i)$-$(iv), we can define a time reflection with respect to the time slice $B_{n_t=0}$ such as $\{r^1,r^2,r^3,r^4\}\rightarrow \{r^1,r^2,r^3,-r^4\}$, so that we can show the reflection positivity as in Refs.~\cite{OS,MM}. But in case of ${\bm w}$ whose some coefficients $w^i$ are larger than or equal to two, for instance $\{w^1,w^2,w^3,w^4\}=\{0,0,1,2\}$, we can not naively define a time reflection with respect to $B_{n_t=0}$: For each lattice site $\{r^1,r^2,r^3,r^4\}$ on $B_{n_t=r^4}$, there exists no corresponding lattice site $\{r^1,r^2,r^3,-r^4\}$ on $B_{n_t=-r^4}$ in general. It seems difficult to show reflection positivity in such bases. 
In numerical calculation in Sec.~\ref{sec:simulation}, we consider only the cases (i)$-$(iv) presented above. 

We also make a comment on the parity transformations 
for the cases of (i)$-$(iv).
Their definitions are given as follows:
\ben
&({\rm i})&\    \{r^1,r^2,r^3,r^4\} \rightarrow \{-r^1,-r^2,-r^3,r^4\},\label{P-i}\\
&({\rm ii})&\   \{r^1,r^2,r^3,r^4\} \rightarrow \{-r^1,-r^2,-r^3,r^4\},\label{P-ii}\\
&({\rm iii})&\  \{r^1,r^2,r^3,r^4\} \rightarrow \{-r^1,-r^3,-r^2,r^4\},\label{P-iii}\\
&({\rm iv})&\   \{r^1,r^2,r^3,r^4\} \rightarrow \{-r^3,-r^2,-r^1,r^4\}.\label{P-iv}
\een
These are compatible with periodicity in the basis $\{\bm E_{\mu}\}$ discussed in \ref{subsec:period}.
The transformations for (i) and (ii) are usual ones. On the other hand, those for (iii) and (iv) involve interchange of coordinates, $r^2\leftrightarrow r^3$ and $r^1\leftrightarrow r^3$, respectively. 

\subsection{Time correlators}
\label{t-corr}
Let us consider time correlators with and without spatial momenta in the general basis $\{\bm E_{\mu}\}$ on the periodic lattice $\Lambda$. 
In the orthonormal basis $\{\bm e_\mu\}$, we take a hadron operator ${\cal O}(n^{\rho})$ at a position $a{\bm n}=an^{\rho}{\bm e_{\rho}}\in\Lambda$. The correlator is given by 
$\langle {\cal O}(n^{\rho})\,{\cal O}(0) \rangle$. Performing the summation over the spatial volume at time $n_t(=n^4)$, the time correlator with zero spatial momentum behaves as 
\ben
&&
a^3
\sum_{n^{\rho}\in B_{n_t}}
\langle {\cal O}(n^{\rho})\,{\cal O}(0) \rangle
\nonumber \\&&\qquad=
C_0\left(e^{-an_tm}+e^{-a(N-n_t)m}\right),
\een
where $C_0$ is an amplitude, $m$ is the hadron mass coupled to ${\cal O}$ and the exited state contributions are omitted. For the correlator with a finite spatial momentum $\vec{p}$, summing over the spatial volume with the weight $\exp(ia\vec{n}\cdot\vec{p})$,  
 the time correlator  behaves as 
\ben
&&
a^3
\sum_{n^{\rho}\in B_{n_t}}
e^{ia\vec{n}\cdot\vec{p}}
\langle {\cal O}(n^{\rho})\,{\cal O}(0) \rangle
\nonumber \\&&\qquad=
C_p\left(e^{-an_t\sqrt{m^2+\vec{p}\cdot\vec{p}}}+e^{-a(N-n_t)\sqrt{m^2+\vec{p}\cdot\vec{p}}}\right),
\een
where $C_{p}$ is an amplitude.

In the basis $\{\bm E_{\mu}\}$, we take ${\cal O}(r^{\rho})={\cal O}(\left(E^{\rho}\right)_{\lambda}n^\lambda)$.\footnote{Since the general basis $\{\bm E_{\mu}\}$ is related by the linear transformation in Eqs.~(\ref{def_E4})-(\ref{detE}) to the orthonormal basis $\{\bm e_\mu\}$, any operator with vector indices such as vector mesons or even any nonlocal operator for staggered fermions  in $\{\bm E_{\mu}\}$ can be always written by a linear combination of the corresponding operators in $\{\bm e_\mu\}$. Thus, the results presented in this subsections can be straightforwardly applied to any  type of fermion discretization as well as any type of operator in general.}  The lattice spacing for ${\bm E_4}$ is given by $a_4=\sqrt{g_{44}}\,a$ in Eq.~(\ref{def_a_mu}), so that the time correlator with zero spatial momentum behaves as
\ben
&&
\frac{a^3}{\sqrt{g_{44}}} 
\sum_{r^{\rho}\in B_{n_t}}
\langle {\cal O}(r^{\rho})\,{\cal O}(0) \rangle
\nonumber \\&&\qquad=
C_0^{\prime}\left(e^{-\sqrt{g_{44}}\,an_tm}+e^{-\sqrt{g_{44}}\,a(N-n_t)m}\right),\label{OO_E}
\een
where $n_t=r^4$, $C_0^{\prime}$ is an amplitude and the excited state contributions are omitted. The factor $\sqrt{g_{44}}$ in the left-hand side of Eq.~(\ref{OO_E})  represents the enlarged spatial volume and it indicates the reduced time interval in the right-hand side. 
Similarly,  the time correlator with a finite momentum $\vec{p}$ behaves as 
\ben
&&
\frac{a^3}{\sqrt{g_{44}}}
\sum_{r^{\rho}\in B_{n_t}}
e^{ia\vec{n}\cdot\vec{p}}
\langle {\cal O}(r^{\rho})\,{\cal O}(0) \rangle
\nonumber \\&&\qquad=
C_p^{\prime}\left(e^{-\sqrt{g_{44}}\,an_t\sqrt{m^2+\vec{p}\cdot\vec{p}}}\right.
\nonumber \\&&\qquad\qquad\quad
\left.+\,\phi(\vec{p})\,e^{-\sqrt{g_{44}}\,a(N-n_t)\sqrt{m^2+\vec{p}\cdot\vec{p}}}\right),\\
&&\phi(\vec{p})=e^{-i\sum_{i=1}^{3}aNp_{i}s^{i}},
\een
where $C_{p}^{\prime}$ is an amplitude and the additional phase factor $\phi(\vec{p})$ comes from  the shift ${aN(s^1\bm E_{1}+s^2\bm E_{2}+s^3\bm E_{3})}$ on $B_{n_t=N}$. Since $p_{i}=2\pi m_{i}/(aN)$ $(m_{i}=0,1,\cdots,N-1)$ the phase factor is given by $\phi(\vec{p})=\exp(-2\pi i\sum_{i=1}^3 m_{i}s^{i})$.



\section{Simulation details}
\label{sec:simulation}

The purpose of numerical calculation is to check the validity of our method so that we employ an experienced simulation setup with less computational cost. Our numerical test is performed in quenched QCD employing the Iwasaki gauge action \cite{iwasaki},
\ben
S
=\frac{\beta}{6}\sum_{\bm n\in\Lambda}\left(c_0\sum_{\mu<\nu} W_{\mu\nu}^{1\times 1}({\bm n})
+c_1\sum_{\mu,\nu} W_{\mu\nu}^{1\times 2}({\bm n}) \right)
\een
with $c_1=-0.331$ for the $1\times 2$ Wilson loop $W_{\mu\nu}^{1\times 2}({\bm n})$ and $c_0=1-8c_1=3.648$ for the $1\times 1$ Wilson loop $W_{\mu\nu}^{1\times 1}({\bm n})$.
We choose $\beta=6/g^2=2.575$, which is one of the bare couplings employed in Ref.~\cite{cp-pacs_2f}. Gauge configurations are generated on a $128^4$ lattice with the periodic boundary condition employing the HMC algorithm in the orthonormal basis $\{\bm e_{\mu}\}$ as usual.  We focus on the energy of the fictitious "$\eta_{\rm ss}$" meson in the various bases as physical quantity, which is easily calculated at a very high level of statistical precision with and without momenta. 
Quark propagators are solved in $\{\bm e_{\mu}\}$ at every 100 trajectories employing the wall source method without gauge fixing \cite{ws_ngf}, and then time correlators of hadrons are constructed in any basis $\{\bm E_{\mu}\}$ including $\{\bm e_{\mu}\}$. All the errors are estimated by a single elimination jackknife method in terms of configurations.

We use an improved Wilson quark action with the mean-field improved clover coefficient $c_{\rm SW}$ defined by
\be
c_{\rm SW}=\left({\overline W}_{\mu\nu}^{1\times 1}\right)^{-3/4}
          =\left(1-0.8412/\beta\right)^{-3/4}=1.345,          
\ee  
where ${\overline W}_{\mu\nu}^{1\times 1}$ is the value in one-loop perturbation theory \cite{cp-pacs_2f}. The lattice spacing is 0.1130(11) fm determined from $m_\rho$ \cite{cp-pacs_2f} so that the spatial extent of the lattice is 14.46(14) fm. One hopping parameter $\kappa=0.1341493786$ 
is chosen to yield $m_{\rm PS}=690.6$ MeV, which corresponds to the "$\eta_{\rm ss}$" meson mass $m_{\eta_{\rm ss}}=\sqrt{2m_K^2-m_\pi^2}=690.6$ MeV. $m_{\rm PS}L$ is more than 50. The simulation parameters are summarized in Table \ref{Tab1}.

In order to investigate cutoff effects on the pseudoscalar meson energies calculated in our method, we generate another quenched QCD ensemble on a $88^4$ lattice at $\beta=2.334$ with the mean-field improved clover coefficient $c_{\rm SW}=1.398$. This lattice has a coarser lattice spacing $0.1632(16)$ fm, where the spatial extent of the lattice is $14.36(14)$ fm. The hopping parameter for "$\eta_{ss}$" meson mass is $\kappa=0.1356019196$.

\begin{table}[t!]
\caption{Simulation parameters. $N_{\rm meas}$ denotes the number of hadron measurements. The lattice spacing $a$ is taken from Ref.~\cite{cp-pacs_2f}.}
\begin{center}
\begin{tabular}{ccccccc}
\hline\hline
$\beta$ & Lattice Size & $a$ [fm] & $c_{\rm SW}$ & $\kappa$ for $m_{\eta_{\rm ss}}$ & $N_{\rm meas}$\\\hline
2.575 & $128^4$ & 0.1130(11) & 1.345 & 0.1341493786 & 120 \\
2.334 & $88^4$  & 0.1632(16) & 1.398 & 0.1356019196 & 120 \\\hline\hline
\end{tabular}
\end{center}
\label{Tab1}
\end{table}%

\section{Results}
\label{sec:results}

\subsection{Effective masses}

\begin{figure}[t!]
	\centering
	\includegraphics[scale=0.33]{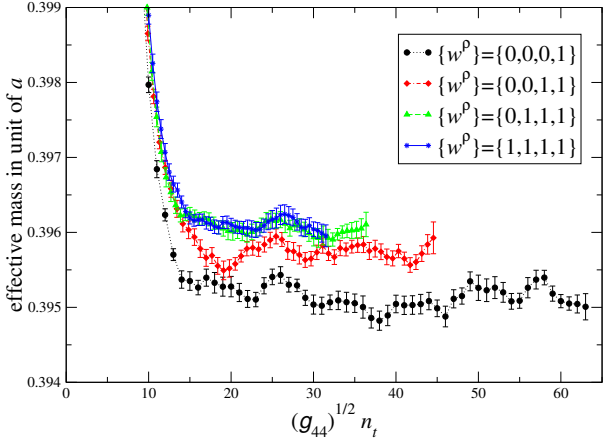}
	\caption{Effective mass of the $"\eta_{\rm ss}"$ meson as a function of $\sqrt{g_{44}}n_t$ along the time direction ${\bm w}$ at $\beta=2.575$. The black filled circle indicates the effective mass for $\{w^\rho\}=\{0,0,0,1\}$ with $\sqrt{g_{44}}=1$, the red filled diamond for $\{w^\rho\}=\{0,0,1,1\}$ with $\sqrt{g_{44}}=1/\sqrt{2}$,  the green filled triangle for $\{w^\rho\}=\{0,1,1,1\}$ with $\sqrt{g_{44}}=1/\sqrt{3}$, and  the blue star for $\{w^\rho\}=\{1,1,1,1\}$ with $\sqrt{g_{44}}=1/2$.} 
	\label{fig2}
\end{figure}

\begin{figure}[t!]
	\centering
	\includegraphics[scale=0.33]{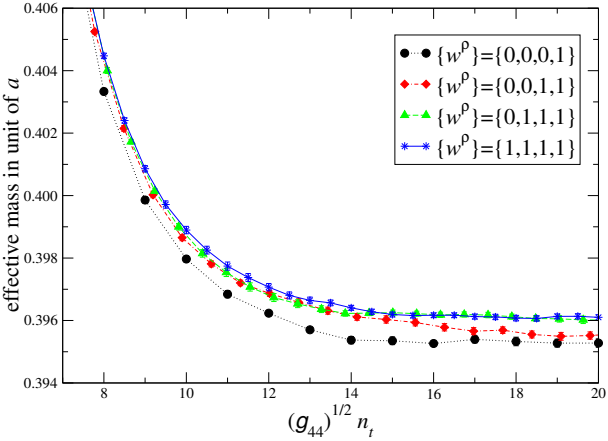}
	\caption{Magnified view of the region of $7\le\sqrt{g_{44}}n_t\le20$ in Fig.\ref{fig2} to clarify the time resolution depending on $\{w^\rho\}$. Error bars are within symbols.}
	\label{fig3}
\end{figure}

\begin{figure}[t!]
	\centering
	\includegraphics[scale=0.33]{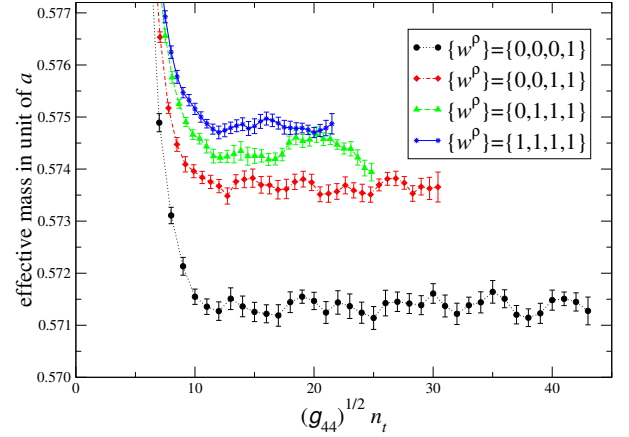}
	\caption{Same as Fig.~\ref{fig2} at $\beta=2.334$.}
	\label{fig4}
\end{figure}

\begin{figure}[t!]
	\centering
	\includegraphics[scale=0.33]{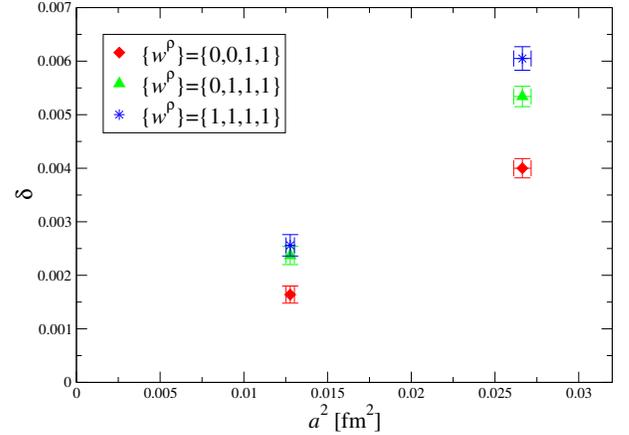}
	\caption{Cutoff dependence of relative mass difference $\delta$ with respect to the  $"\eta_{\rm ss}"$ meson mass along $\{w^\rho\}=\{0,0,0,1\}$. 
The red filled diamond represents $\delta$ for $\{w^\rho\}=\{0,0,1,1\}$, the green filled triangle  for $\{w^\rho\}=\{0,1,1,1\}$, and the blue star for $\{w^\rho\}=\{1,1,1,1\}$.}
	\label{fig5}
\end{figure}

In Fig.~\ref{fig2} we first show the effective mass of the $"\eta_{\rm ss}"$ meson as a function of $\sqrt{g_{44}}n_t$ with four choices of the time direction ${\bm w}$ at $\beta=2.575$. Figure~\ref{fig3} is a magnified version of Fig.~\ref{fig2} choosing the time interval of $7\le\sqrt{g_{44}}n_t\le20$. We observe that the lattice spacing in the time direction is varied according to ${\bm w}$. The case of $\{w^\rho\}\equiv\{w^1,w^2,w^3,w^4\}=\{1,1,1,1\}$ gives the finest lattice spacing, which is reduced to a half as compared to the case of $\{w^\rho\}=\{0,0,0,1\}$. 
Numerical values of the fit results for the $"\eta_{\rm ss}"$ meson mass are listed in Table~\ref{fit_mass_b2575}. We find clear deviation beyond the error bar among four cases of ${\bm w}$: The biggest is $0.25$\% difference between the cases of $\{w^{\rho}\}=\{0,0,0,1\}$ and $\{1,1,1,1\}$, each of which yields less than $0.017$\% error for the $"\eta_{\rm ss}"$ meson mass. The situation is quantitatively clarified by defining the relative mass difference $\delta=m_{\rm PS}|_{\{w^\rho\}}/m_{\rm PS}|_{\{0,0,0,1\}}-1$, which is also presented in Table~\ref{fit_mass_b2575}.  This small but solid difference could be a cutoff effect due to the change of lattice geometry. In order to check our speculation, we have repeated the calculation on a coarser lattice at $\beta=2.334$ with the same physical volume and the $"\eta_{\rm ss}"$ meson mass as in the case of $\beta=2.575$. The results for the $"\eta_{\rm ss}"$ meson effective mass are plotted in Fig.~\ref{fig4}. Table~\ref{fit_mass_b2334} summarizes the numerical values for the results of the $"\eta_{\rm ss}"$ meson mass and their relative differences. We observe the similar deviation among four cases of ${\bm w}$ as in the case of $\beta=2.575$. In Fig.~\ref{fig5}, we plot the relative mass difference $\delta$ at $\beta=2.575$ and $2.334$ as a function of $a^2$. As the lattice spacing diminishes, the mass difference monotonically reduces as we expected.

\begin{table}[t!]
\caption{Fit masses for each time direction ${\bm w} = w^\rho {\bm e_\rho}$ at $\beta=2.575$ on a $128^4$ lattice. The fit range is $\sqrt{g_{44}}n_t\ge 20$. $\delta$ is the relative mass difference with respect to the mass for $\{w^\rho\}=\{0,0,0,1\}$.}
\begin{center}
\begin{tabular}{c|cccc}
\hline\hline
$\{w^\rho\}$ & $\{0,0,0,1\}$& $\{0,0,1,1\}$& $\{0,1,1,1\}$& $\{1,1,1,1\}$\\\hline
$am_{\rm PS}$ & 0.395107(46)&0.395756(45)&0.396044(50)&0.396120(65) \\
$\delta$ & 0 & +0.00164(16)& +0.00237(17) & +0.00256(20)\\
\hline\hline
\end{tabular}
\end{center}
\label{fit_mass_b2575}
\end{table}%

\begin{table}[t!]
\caption{Fit masses for each time direction ${\bm w} = w^\rho {\bm e_\rho}$ at $\beta=2.334$ on a $88^4$ lattice. The fit range is $\sqrt{g_{44}}n_t\ge 14$. $\delta$ is the relative mass difference with respect to the mass for $\{w^\rho\}=\{0,0,0,1\}$.}
\begin{center}
\begin{tabular}{c|cccc}
\hline\hline
$\{w^\rho\}$ & $\{0,0,0,1\}$& $\{0,0,1,1\}$& $\{0,1,1,1\}$& $\{1,1,1,1\}$\\\hline
$am_{\rm PS}$ &0.571371(72)&0.573658(77)&0.574420(63)&0.574830(78)  \\
$\delta$ & 0 & +0.00400(18) & +0.00534(19) & +0.00605(22)\\
\hline\hline
\end{tabular}
\end{center}
\label{fit_mass_b2334}
\end{table}%

To investigate a possible origin of the mass difference among the bases, let us consider a dispersion relation of a free scaler field with zero spatial momentum in the general basis $\{\bm E_{\mu}\}$, which is given by
\begin{eqnarray}
\frac{4}{g_{44}} \sinh^2\left(\frac{\sqrt{g_{44}}\,a m_{\rm PS}}{2}\right) &=& (am_{\eta_{\rm ss}})^2,
\label{dispersion_zero_p}
\end{eqnarray}
where $m_{\rm PS}$ is the physical mass, namely the pole mass, and $m_{\eta_{\rm ss}}$ is the bare mass. In the continuum limit, $m_{\rm PS}=m_{\eta_{\rm ss}}$.  To evaluate the cutoff  effects on $m_{\rm PS}$ we expand $m_{\rm PS}$ in terms of the lattice spacing $a$:
\begin{eqnarray}
m_{\rm PS} &=& m_{\eta_{\rm ss}}+ a m^{(1)}_{\rm PS} + a^2 m^{(2)}_{\rm PS} + O(a^3).
\label{m_expansion}
\end{eqnarray}
Solving Eqs.~(\ref{dispersion_zero_p}) and (\ref{m_expansion}) up to $O(a^2)$ we find that the $O(a)$ term vanishes, $m^{(1)}_{\rm PS}=0$, and the $O(a^2)$ term is given by 
\begin{eqnarray}
m^{(2)}_{\rm PS} &=& -\frac{g_{44}}{24}m_{\eta_{\rm ss}}^3.
\end{eqnarray}
So the relative mass difference $\delta$ up to $O(a^2)$ is expressed as
\begin{eqnarray}
\delta &=& \frac{1}{24}a^2 m_{\eta_{\rm ss}}^2 (1-g_{44}).
\end{eqnarray}
At $\beta=2.575$ this expression yields  $\delta=0.0033$, $0.0044$ and $0.0049$ for $\{w^{\rho}\}=\{0,0,1,1\}, \{0,1,1,1\}$ and $\{1,1,1,1\}$, and at $\beta=2.334$ we obtain $\delta=0.0068, 0.0091$ and $0.0102$ for $\{w^{\rho}\}=\{0,0,1,1\},\ \{0,1,1,1\}$ and $\{1,1,1,1\}$. We find that the measured relative mass difference $\delta$ can be qualitatively explained by the $O(a^2)$ effects on the pseudoscaler meson mass.

\subsection{Effective energies with finite momenta}

\begin{figure}[t!]
	\centering
	\includegraphics[scale=0.33]{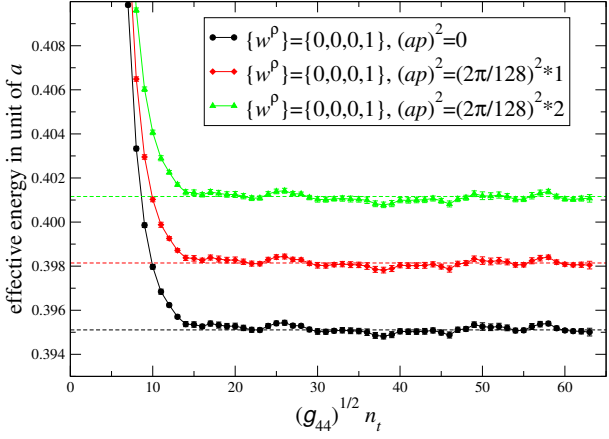}
	\caption{Effective energies with momenta at $\beta=2.575$ as a function of $\sqrt{g_{44}}n_t=n_t$ choosing $\{w^\rho\}=\{0,0,0,1\}$ for the time direction. 
	 The black filled circle, the red filled diamond and the green filed triangle are the effective energies with spatial momentum squared $(ap)^2=0, (2\pi/128)^2\times1$ and $(2\pi/128)^2\times 2$, respectively. The black dashed line represents the central value of the fit mass $am_{\rm PS}$ in Table \ref{fit_mass_b2575}. The red and green dashed lines are the expected energies given by $\sqrt{(am_{\rm PS})^2+(ap)^2}$ with spatial momentum squared $(ap)^2=(2\pi/128)^2\times1$ and $(2\pi/128)^2\times 2$, respectively.}
	\label{fig6}
\end{figure}

\begin{figure}[t!]
	\centering
	\includegraphics[scale=0.33]{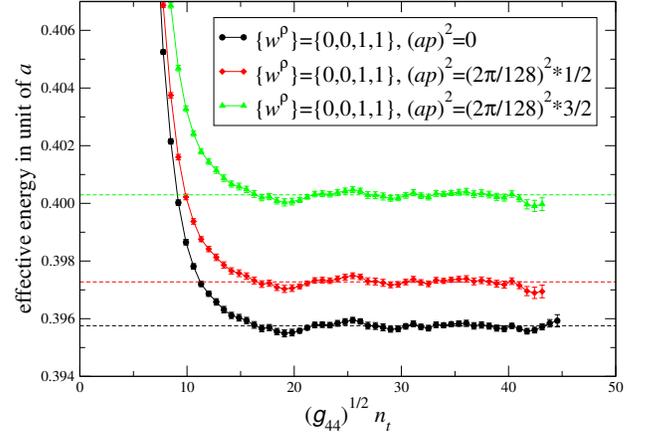}	
	\caption{Same as Fig.~\ref{fig6} for $\{w^\rho\}=\{0,0,1,1\}$. In this case, $\sqrt{g_{44}}n_t=n_t/\sqrt{2}$. The red and green dashed lines are the expected energies with spatial momentum squared $(ap)^2=(2\pi/128)^2\times1/2$ and $(2\pi/128)^2\times 3/2$, respectively.}
	\label{fig7}
\end{figure}

\begin{figure}[t!]
	\centering
	\includegraphics[scale=0.33]{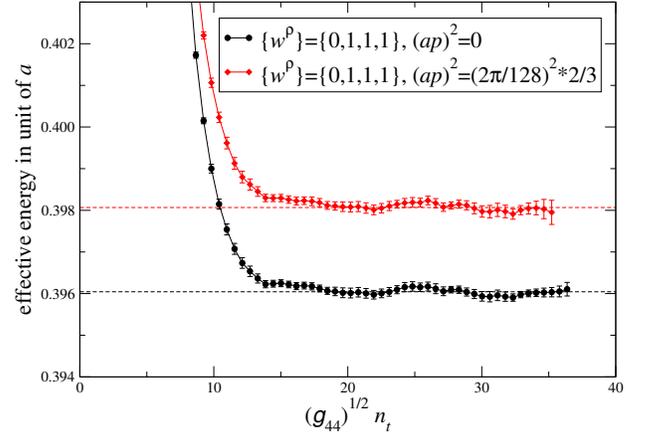}	
	\caption{Same as Fig.~\ref{fig6} for $\{w^\rho\}=\{0,1,1,1\}$. In this case, $\sqrt{g_{44}}n_t=n_t/\sqrt{3}$. The red dashed line is the expected energy with spatial momentum squared $(ap)^2=(2\pi/128)^2\times2/3$.}
	\label{fig8}
\end{figure}

\begin{figure}[t!]
	\centering
	\includegraphics[scale=0.33]{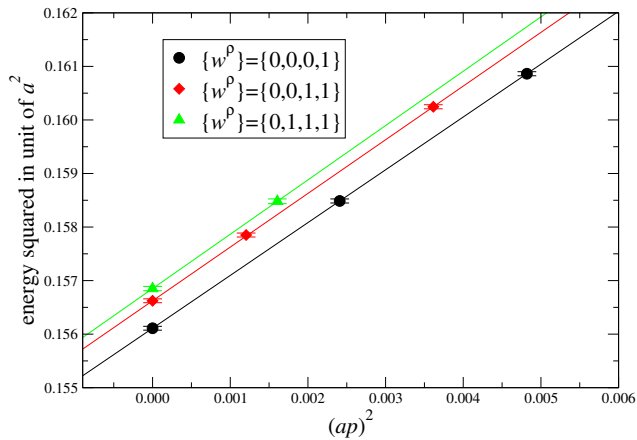}
	\caption{Dispersion relation of $"\eta_{\rm ss}"$ meson at $\beta=2.575$. The data correspond to those in Figs.~\ref{fig6}, \ref{fig7} and \ref{fig8}. The black filled circle represents the dispersion relation for $\{w^\rho\}=\{0,0,0,1\}$, the red filled diamond  for $\{w^\rho\}=\{0,0,1,1\}$ and the green triangle for $\{w^\rho\}=\{0,1,1,1\}$. Solid lines are for eye-guide.}
	\label{fig9}
\end{figure}

\begin{table}[t!]
\caption{Fit energies with momenta for each time direction ${\bm w} = w^\rho {\bm e_\rho}$ at $\beta=2.575$ on a $128^4$ lattice. $aE$ is the fit energy in unit of $a$. 
$p^2/(2\pi/128a)^2$ is the spatial momentum squared in unit of $(128a/2\pi)^2$.  
The fit range is $\sqrt{g_{44}}n_t\ge 20$. }
\begin{center}
\begin{tabular}{ccc}
\hline\hline
$\{w^\rho\}$ & $ p^2/(2\pi/128a)^2$ & $aE$ \\\hline
$\{0,0,0,1\}$ & $0$ & 0.395107(46)\\
$\{0,0,0,1\}$ & $1$ & 0.398104(47)\\
$\{0,0,0,1\}$ & $2$ & 0.401077(47)\\\hline
$\{0,0,1,1\}$ & $0$ & 0.395756(45)\\
$\{0,0,1,1\}$ & $1/2$ & 0.397305(47)\\
$\{0,0,1,1\}$ & $3/2$ & 0.400307(47)\\\hline
$\{0,1,1,1\}$ & $0$ & 0.396044(50)\\
$\{0,1,1,1\}$ & $2/3$ & 0.398097(55)\\\hline\hline
\end{tabular}
\end{center}
\label{dispersion}
\end{table}%

Now let us turn to the $"\eta_{\rm ss}"$ meson energies with finite momenta.
In Fig.~\ref{fig6}, we plot the effective energies with spatial momentum squared $(ap)^2=0$ (black filled circle), $(2\pi/128)^2\times 1$ (red filled diamond) and $(2\pi/128)^2\times 2$ (green filled triangle) along the time direction of $\{w^\rho\}=\{0,0,0,1\}$ at $\beta=2.575$. The black dashed line denotes the value of $am_{\rm PS}$ in Table \ref{fit_mass_b2575}, which is the $"\eta_{\rm ss}"$ meson mass, and the red and green dashed lines are for the expected energies estimated by $\sqrt{(am_{\rm PS})^2+(ap)^2}$. We observe that the measured effective energies with finite momenta show good consistencies with the expected ones. Figure~\ref{fig7} shows effective energies along the time direction of $\{w^\rho\}=\{0,0,1,1\}$. In this case, the minimum and the second minimum fractional spatial momentum squared in unit of $(2\pi/128)^2$ are $(2\pi/128)^2\times 1/2$ (red filled diamond) and $(2\pi/128)^2\times 3/2$ (green filled triangle)\footnote{The second minimum momentum squared is $(2\pi/128)^2$, which is not fractional in unit of $(2\pi/128)^2$ and out of our interest here.}. We find that the $"\eta_{\rm ss}"$ meson successfully acquires the expected energies with finite momenta. Figure~\ref{fig8} is for the case of $\{w^\rho\}=\{0,1,1,1\}$, whose minimum spatial momentum squared is $(2\pi/128)^2\times 2/3$ (red filled diamond). Again we find good consistency between the measured effective energy and the expected one. In Fig.~\ref{fig9}, we present the dispersion relation for the $"\eta_{\rm ss}"$ meson, where the numerical values of the fit energies are summarized in Table~\ref{dispersion}. We find that the red and green symbols slightly shift upward from the black ones. This could be due to the cutoff effects discussed in the previous subsection.

\section{Conclusions and outlook}
\label{sec:conclusion}

We have proposed an efficient use of geometry in lattice QCD simulations. An appropriate reconstruction of a given periodic lattice provides us an opportunity to improve the resolution of the spatial momentum and the time interval. We have explained the theoretical details and presented some numerical results to demonstrate the validity of our method. We have also found that tiny cutoff effects depending on the time direction ${\bm w}$ safely diminish toward the continuum limit.
The reconstruction method could help us investigate the hadron structures, the excited states, the heavy hadrons etc., especially, in the master-field simulations.

\begin{acknowledgments}
The authors thank the members of the PACS Collaboration for helpful discussions and comments.  
Numerical calculations are carried out on Oakforest-PACS through the HPCI System Research project (Project ID: hp170093 , hp180051) and the Interdisciplinary Computational Science Program in the Center for Computational Sciences, University of Tsukuba.
This work is supported in part by Grants-in-Aid for Scientific Research from the Japan Society for the Promotion of Science (JSPS) (No. 16K13798).
\end{acknowledgments}


\end{document}